\documentclass[fleqn,showpacs,superscriptaddress,nofootinbib]{revtex4}

\usepackage{amsmath}
\usepackage{graphicx,psfrag}

\def\slash#1{\setbox0=\hbox{$#1$}               % set a box for #1
        \dimen0=\wd0                            % and get its size
        \setbox1=\hbox{/} \dimen1=\wd1          % get size of /
        \ifdim\dimen0>\dimen1                   % #1 is bigger
        \rlap{\hbox to \dimen0{\hfil/\hfil}}    % so center / in box
        #1                                      % and print #1
        \else                                   % / is bigger
        \rlap{\hbox to \dimen1{\hfil$#1$\hfil}} % so center #1
        /                                       % and print /
        \fi}                                    %

\begin{document}

\title{Spin asymmetries in jet-hyperon production at LHC}

\author{D. Boer}
\email{D.Boer@few.vu.nl}
\affiliation{
Department of Physics and Astronomy, Vrije Universiteit Amsterdam,\\
NL-1081 HV Amsterdam, the Netherlands}

\author{C.J. Bomhof}
\email{cbomhof@nat.vu.nl}
\affiliation{
Department of Physics and Astronomy, Vrije Universiteit Amsterdam,\\
NL-1081 HV Amsterdam, the Netherlands}

\author{D.S. Hwang}
\email{dshwang@sejong.ac.kr}
\affiliation{Department of Physics, Sejong University,\\
Seoul 143--747, South Korea}

\author{P.J. Mulders}
\email{mulders@few.vu.nl}
\affiliation{
Department of Physics and Astronomy, Vrije Universiteit Amsterdam,\\
NL-1081 HV Amsterdam, the Netherlands}

\begin{abstract}
We consider polarized $\Lambda$ hyperon production in
proton-proton scattering, $p \, p \to \left(
\Lambda^\uparrow \text{jet}\right) \, \text{jet} \, X$, 
in the kinematical region of the LHC experiments, in particular the ALICE
experiment. 
We present a new $\Lambda$ polarization observable that arises from the Sivers
effect in the fragmentation process. It can be large even at 
midrapidity and therefore, is of interest for high energy hadron 
collider 
experiments. Apart from its potential to shed light on the mechanisms behind 
the phenomenon of $\Lambda$ polarization arising in unpolarized hadronic
collisions, the proposed 
observable in principle also allows to test the possible 
color flow dependence of single spin asymmetries and the (non)universality of 
transverse momentum dependent fragmentation functions.  
\end{abstract}
\date{\today}

\maketitle

\section{Introduction}

Since the observation of large transverse polarization of produced $\Lambda$
hyperons in the inclusive reactions $pp\to \Lambda^{\uparrow}X$ \cite{lesnik75}
and $p\ Be\to \Lambda^{\uparrow}X$ \cite{bunce76} in the middle of the 1970's,
there have been many experimental and
theoretical investigations aimed at understanding this striking polarization
phenomenon~\cite{expts, theors}. The polarization measurements of
$\Lambda$ hyperons produced in these inclusive
reactions have been performed in fixed target experiments, and the data
showed that the $\Lambda$ polarizations are large only for large $x_F$.
This poses a problem if one wants to investigate this observable
further using high energy colliders such as RHIC, Tevatron or LHC, where the
capabilities to measure $\Lambda$ polarization are restricted to the
midrapidity region, where $x_F$ is very small. However, in this
paper we point out a way to by-pass this problem by making a
less inclusive measurement: to select two-jet events and to measure the
jet momenta $K_j^{}$ and $K_{j'}^{}$ in addition to the momentum
$K_\Lambda^{}$ and polarization $S_\Lambda^{}$ of the $\Lambda$ that is part of
either of the two jets. An asymmetry proportional to 
$\epsilon_{\mu\nu\alpha\beta} K_j^{\mu} K_{j'}^{\nu}
K_\Lambda^{\alpha} S_\Lambda^\beta$ can then arise, which is neither
power suppressed, nor needs to be zero\footnote{In $pp\to \Lambda^{\uparrow}X$ the $\Lambda$ polarization 
needs to vanish at $x_F=0$ due to symmetry reasons. In $p\, A\to
\Lambda^{\uparrow}X$ nuclear effects 
could allow for a nonvanishing asymmetry at midrapidity. No such effects have
been observed thus far.} at vanishing $x_F$ of the
$\Lambda$. The observable is proportional to 
$(\boldsymbol K_j{\times}\boldsymbol K_\Lambda){\cdot}\boldsymbol S_\Lambda$
in the center of mass frame of the two jets.
Within a factorized description of the process under consideration, 
such an asymmetry can arise from spin and transverse momentum
dependent fragmentation functions, which in turn can arise from two types of
interactions. We will point out how the measurement of the suggested 
asymmetry in principle can allow for a differentiation between these two
types of mechanisms. In this way also hadron collider experiments can
contribute to our understanding of the underlying mechanisms that lead to
$\Lambda$ polarization, 
in addition to experiments with electron-positron and 
lepton-hadron collisions. 

\begin{figure}[htb]
\centering
\psfrag{f}[cc][cc]{{\large $f$}}
\psfrag{T}[lc][lc]{$\,T$}
\psfrag{k}[lc][lc]{{\large$k$}}
\psfrag{q}[cc][cc]{{\large$q$}}
\psfrag{P}[cc][cc]{{\large$P$}}
\psfrag{s}[cc][cc]{{\large$\,S$}}
\psfrag{1T}[cc][cc]{$1T$}
\includegraphics[width=8cm]{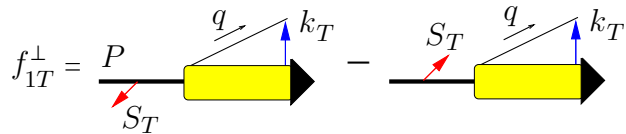}\nopagebreak\\
\parbox{0.8\textwidth}{\caption{Schematic depiction of the 
Sivers distribution function. The spin vector $S_T^{}$ points out of and into
the page, respectively. \label{SiversDistribution}}}
\end{figure}

It is well-known that transverse momentum dependent distribution and 
fragmentation functions, nowadays commonly referred to as TMDs, can have a 
nontrivial spin dependence and that the so-called ``$T$-odd'' TMDs can lead to 
single spin asymmetries~\cite{Sivers,Collins:1992kk,Anselmino:1994tv,BM98}.
They are also often referred to as ``naively $T$-odd'',
because the appearance of these functions does not imply a violation of
time-reversal invariance. The Sivers distribution function $f_{1T}^\perp$, 
schematically depicted in Fig.\
\ref{SiversDistribution}, is the oldest example of such functions.  
It describes the difference between the momentum distributions of quarks 
inside protons transversely polarized in opposite directions. 
The Sivers effect was put forward 
\cite{Sivers,Anselmino:1994tv} as a possible explanation for the large single 
spin asymmetries observed in $p^\uparrow \, p \to \pi \, X$ experiments 
\cite{Adams}. Furthermore, it generates single spin asymmetries in
semi-inclusive DIS \cite{BM98,Brodsky:2002cx},
which have also been measured to be nonzero 
\cite{Airapetian:2004tw}, and it results e.g.\ 
in asymmetric di-jet correlations in 
$p^\uparrow \, p \to \text{jet} \, \text{jet}\, X$
\cite{Boer:2003tx,Bacchetta:2005rm}, which however are 
not yet visible in the data analyzed \cite{Abelev:2007ii}. 

\begin{figure}[htb]
\centering
\psfrag{D}[cc][cc]{{\large$D$}}
\psfrag{T}[cc][cc]{$\,T$}
\psfrag{k}[cc][cc]{{\large$k$}}
\psfrag{S}[rc][rc]{{\large$S$}}
\psfrag{1T}[cc][cc]{$1T$}
\includegraphics[width=8cm]{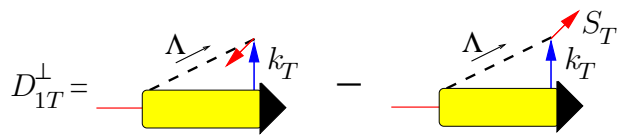}\\
\parbox{0.8\textwidth}{\caption{The 
Sivers or ``polarizing'' 
fragmentation function.\label{SiversFragmentation}}}
\end{figure}

\begin{figure}
\centering
\begin{minipage}[t]{6cm}
\centering
final-state interactions:\\[2mm]
\includegraphics[width=0.45\textwidth]{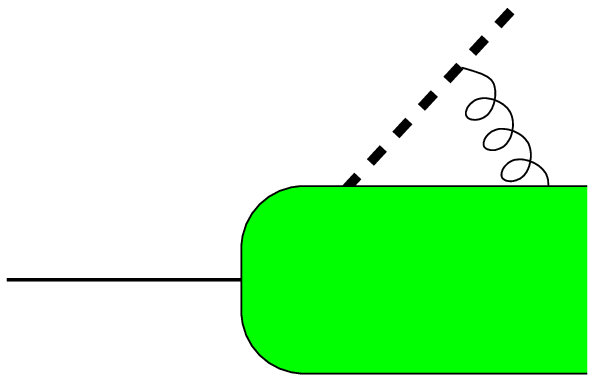}
\put(-39,0){\vector(0,-1){18}}\\[1mm]
$D_{1T}^\perp$\\[2mm]
(a)
\end{minipage}\hspace{0.7cm}
\begin{minipage}[t]{6cm}
\centering
soft gluon interactions (gauge links):\\[2mm]
\includegraphics[width=0.45\textwidth]{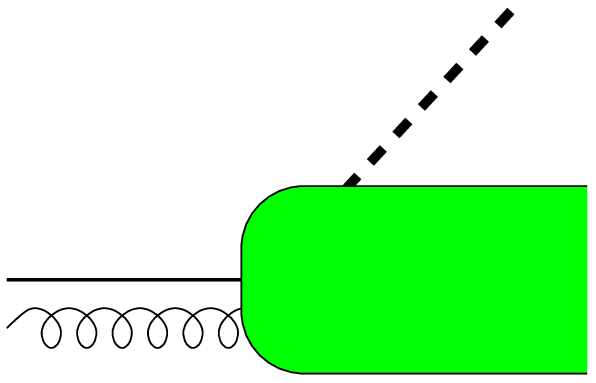}
\put(-39,0){\vector(0,-1){18}}\\[1mm]
$\widetilde D{}_{1T}^\perp$\\[2mm]
(b)
\end{minipage}
\parbox{0.8\textwidth}{\caption{
Two possible mechanisms to generate single spin asymmetries in the fragmentation process:
(a) through final-state interactions within the out-state composed of 
the outgoing hadron and the rest of the jet;
(b) through soft gluonic interactions between the jet and the hard part.
\label{ToddFragmentation}}}
\end{figure}

The fragmentation analogue of the Sivers distribution function is called
$D_{1T}^\perp$ \cite{Mulders:1995dh}. It describes the distribution of 
transversely polarized spin-1/2 hadrons, such as $\Lambda$'s, inside the
jet of a fragmenting {\em unpolarized\/} quark, 
cf.\ Fig.\ \ref{SiversFragmentation}. For this reason it has been 
referred to as ``polarizing fragmentation function'' in Ref.~\cite{ABDM01}.
It is an odd function of the transverse
momentum of the observed hadron w.r.t.\ the quark direction, or equivalently, 
the jet direction. 
Despite the similarity between the definitions of $D_{1T}^\perp$ and 
$f_{1T}^\perp$, there are some important differences. 
Nonvanishing $T$-odd distribution functions require soft gluonic interactions 
between the target remnants and the active partons~\cite{Brodsky:2002cx}.
These interactions can be resummed into Wilson lines (gauge links),
ensuring the gauge invariance of the operator definitions of the 
distribution
functions~\cite{Collins:2002kn,JiYuan,Belitsky:2002sm,Boer:2003cm}. 
On the other hand, there are two mechanisms to 
generate $T$-odd effects in the fragmentation process: through final-state 
interactions within the jet, e.g.\ between the observed outgoing 
hadron and the rest of the jet~\cite{Collins:1992kk}, or through soft gluonic 
interactions between the jet and the hard scattering part 
(sometimes also referred to as final-state interactions), 
see Figs~\ref{ToddFragmentation}a and \ref{ToddFragmentation}b respectively.
As for the distribution functions,
the latter interactions give rise to Wilson lines.
Both effects can be expressed in terms of a TMD fragmentation function.
The $\Lambda$ polarization observable that is the subject of this paper 
in principle allows for a differentiation between these two effects. 
There is considerable interest in this issue, as it 
could shed light on the color flow dependence of single spin
asymmetries and the (non)universality of transverse momentum dependent
fragmentation functions. 

In recent years it has become apparent that $T$-odd TMDs enter in
different ways in different processes, depending on the color flow in
the partonic subprocesses\footnote{Although the possibility of such 
effects for $T$-even TMDs are not excluded, they will not be considered 
here.}. 
The sign relation between the Sivers function appearing 
in semi-inclusive DIS and in Drell-Yan scattering was the first example of such
process dependence \cite{Collins:2002kn,BroHwaSch}. 
More complicated relations were discussed
soon afterwards for hadron production processes in hadronic collisions
\cite{Bomhof:2004aw,Bomhof:2006dp,Bomhof:2006ra,Ratcliffe:2007ye}. 
Process dependence may seem at odds with factorization \cite{Collins:2007nk}, 
but since the color flow dependence can be explicitly taken into account
through the determination of the Wilson line structure,
the process dependence is explicitly calculable and 
hence, predictive power may be 
retained albeit in a less straightforward manner. For the specific example of 
the single spin asymmetry in 
$p^\uparrow \, p \to \text{jet} \, \text{jet}\, X$ the effects of color flow
dependence have been taken into account \cite{Bomhof:2007su}, 
leading to a suppression w.r.t.\ the standard parton model expectation. 
For fragmentation 
functions the situation is complicated further by the fact that only one of
the two mechanisms leading to nonzero $T$-odd fragmentation functions depends 
on the color flow structure \cite{Boer:2003cm}. 
However, a claim based on a model calculation \cite{Metz:2002iz} 
and on more general arguments \cite{Collins:2004nx} has been put forward 
that for $T$-odd fragmentation functions no color flow dependence arises at 
all. This makes an experimental test of this dependence, or of its
absence, all the more interesting.

We propose to measure it in the process $p \, p \to \left(
\Lambda^\uparrow \text{jet}\right) \, \text{jet} \, X$, where the brackets 
indicate that the $\Lambda$ is part of one of the two observed jets which 
are almost back-to-back in the plane perpendicular to the beam axis and its
polarization is measured.
This measurement can be done for instance at LHC and RHIC, 
although --as we will point out--
the situation at LHC is more straightforward thanks to the dominance of 
gluon-gluon scattering in the region of interest. 

The $\Lambda$
polarization observable we put forward here is similar to the Sivers
effect observable in $p^\uparrow \, p \to \text{jet} \, \text{jet}\,
X$~\cite{Boer:2003tx,Bacchetta:2005rm}, 
but the $\Lambda$ polarization observable does not depend on a
non-collinearity of jets in the transverse plane. 
In other words, even if one were to assume the
initial partons to be collinear to the parent hadrons, in which case the jets
are exactly back-to-back in the plane perpendicular to the beam axis, an
asymmetry can still arise. Therefore, it is expected that the 
$\Lambda$ polarization
observable is less sensitive to effects that would smear out the partonic $2
\to 2$ kinematics. This prompts us to ignore contributions that depend on
nontrivial spin and transverse momentum dependent effects in the initial
states (which would be proportional to $h_1^\perp H_1$ in the notation of
Ref.\ \cite{BM98}).   

The observable turns out to be 
very sensitive to the unpolarized $\Lambda$
fragmentation functions. The relative importance of contributing
partonic subprocesses differs strongly
depending on whether one uses for instance the fragmentation functions by 
De Florian et al.~\cite{FSV} (FSV) or by Albino et al.~\cite{AKK} (AKK). 
Therefore, 
even if no polarization effect is seen, the unpolarized data obtained by the
proposed measurement
can in any case lead to a considerable improvement in the determination of the 
unpolarized $\Lambda$ fragmentation functions.   

We will explain why it is necessary to measure both jet momenta and also how
it differs from the process $p \, p \to \Lambda^\uparrow \, X$. We will start
with a discussion of that process. Anselmino et al.~\cite{ABDM01} 
have attempted to describe the latter in terms of
$D_{1T}^\perp$, but although a satisfactory description of the $\Lambda$
polarization at larger values of the transverse momentum of the
$\Lambda$ could be obtained,
considerable doubts about the applicability of the assumed factorization have
arisen afterwards. A similar problem is unlikely to arise for the $\Lambda$
polarization study in the process $p \, p \to \left(
\Lambda^\uparrow \text{jet}\right) \, \text{jet} \, X$ at LHC and RHIC.

\section{$\Lambda$
polarization  in  $p \, p \to \Lambda^\uparrow \, X$}

The $\Lambda$'s measured in the process 
$p \, p \to \Lambda^\uparrow \, X$~\cite{lesnik75} 
by the fixed target experiments have
transverse momenta $p_T^{}$ w.r.t.\ the beam direction ranging up to 4 GeV$/c$. 
The $\Lambda$ polarization transverse to the production
plane first grows linearly with $p_T^{}$, but around $p_T^{} \approx 1$
GeV$/c$ it levels off and remains approximately constant. 
For large $p_T^{}$ one expects
the polarization to fall off as $1/p_T^{}$, based on theoretical (collinear
expansion) arguments \cite{Kane:1978nd}. The high energy
collider experiments would be ideally suited to demonstrate this
power-law fall-off behavior, if it were not for the small $x_F$ values
probed in those experiments. 

The linearly rising behavior at very low $p_T$ is well-described by
recombination models employing spin-orbit couplings \cite{theors}. 
However, the plateau region for $p_T^{} \sim 1-4$ GeV$/c$ is
less straightforward to incorporate in such soft physics models.    
Anselmino et al.~\cite{ABDM01} assumed that the $p_T^{}$ in this region
may be considered a hard scale which then would allow for factorization of 
the cross section expression. However, since the scale is not too
large either, instead of restricting to collinear factorization, 
transverse momentum dependence at the parton level was included. In this way
the $\Lambda$ polarization could be described in terms of the 
fragmentation function $D_{1T}^\perp$. Although a 
satisfactory description of the higher-$p_T^{}$ data was obtained in
this way, a preliminary study of the unpolarized cross section within the same
factorized approach resulted in considerable disagreement with the
data, as mentioned in Ref.\ \cite{ABDM02}. 
Apparently, at large $x_F$ a $p_T^{}$ value of 3 or 4 GeV$/c$ is not yet 
large enough to
allow for a factorized description in the case of $\Lambda$ production, 
as opposed to for instance pion production. Inclusion of
intrinsic transverse momentum in the initial state 
parton distributions may lead to
less discrepancy with the data, as studied for pion production 
\cite{Anselmino:2004ky}, but remains to be done for $\Lambda$
production. 

Despite the doubts about whether $D_{1T}^\perp$ can be extracted from 
the existing fixed-target $\Lambda$ polarization data, the collider 
experiments at RHIC and LHC are in a kinematic region that
should allow for a factorized description of the cross section and
thus can be considered safe. As mentioned, they do have the drawback that
$\Lambda$ measurements are generally restricted to small values of $x_F$,
where the $\Lambda$ polarization in $p \, p \to \Lambda^\uparrow \, X$ will be
close to zero. Therefore, we next discuss a different process, where 
this drawback does not appear. 

The main difference between the traditional process 
$p \, p \to \Lambda^\uparrow \, X$ and the one considered in this paper, 
$p \, p \to \left(\Lambda^\uparrow \text{jet}\right) \, \text{jet} \, X$ comes
from the way $D_{1T}^\perp$ enters in the asymmetry expressions. In the latter
process it enters in an unintegrated way, i.e.\ the 
asymmetry is proportional to $D_{1T}^\perp(z,k_T^2)$, where $z$ and $k_T$ are
observed. In the traditional process on the other hand, there is an 
integration over $k_T$, which implies that one essentially probes the first
derivative of the partonic cross section (in case of collinear initial
partons), which results in the aforementioned $1/p_T$ dependence. In 
$p \, p \to \left(\Lambda^\uparrow \text{jet}\right) \, \text{jet} \, X$ 
this $1/p_T$ factor does not arise.

\section{$\Lambda$
polarization  in  $p \, p \to \left(
\Lambda^\uparrow \text{jet}\right) \, \text{jet} \, X$}

We consider the process 
$p(P_1){+}p(P_2)
{\rightarrow}\left(\Lambda^\uparrow(K_\Lambda)\text{jet}(K_j)\right)
{+}\text{jet}(K_{j'}){+}X$.
As indicated by the brackets, the $\Lambda$ is part of the jet with 
observed momentum $K_j$,
where the two jets are almost back-to-back in the plane perpendicular 
to the beam axis.
The projections of the outgoing momenta in this plane will be denoted by
$K_{\Lambda\perp}$, $K_{j\perp}$ and $K_{j'\perp}$.
We will also use the scaled variables 
$x_{j\perp}\,{=}\,2|\boldsymbol K_{j\perp}|/\sqrt s$ and
$x_{j'\perp}\,{=}\,2|\boldsymbol K_{j'\perp}|/\sqrt s$.
At leading order the hadronic process is mediated by two-to-two partonic
scattering $a(p_1)b(p_2){\rightarrow}c(k)d(k')$. 
Since in the process under consideration both jet momenta are observed,
the momenta of the outgoing partons are known,
i.e.\ $k\,{\equiv}\,K_j$ and $k'\,{\equiv}\,K_{j'}$.
For the parton momenta it is advantageous to make the Sudakov decompositions
$p_i\,{=}\,x_iP_i{+}\sigma_in_i{+}p_{iT}$ for incoming partons and 
$k\,{=}\,\frac{1}{z}K_\Lambda{+}\sigma n{+}k_T$ for the outgoing parton
that will fragment into the hyperon. 
The $n$-vectors are lightlike four-vectors that have nonzero overlaps with the
associated hadron momentum (i.e.\ $n_i{\cdot}P_i\,{\neq}\,0$ and
$n{\cdot}K_\Lambda\,{\neq}\,0)$. There is a certain degree of 
arbitrariness in the choice of these $n$-vectors, but it is required that 
$\sigma_i$ and $\sigma$ are suppressed by inverse powers of one of the 
hard scales in the process (such as $\sqrt{s}$ or $|\boldsymbol K_{j\perp}|$).
This requirement ensures in the first place that the cross section can be
expressed in terms of TMDs, i.e.\ functions of only the 
momentum fractions and intrinsic transverse momenta. 

By taking $n$ to be a measurable four-vector in the process, e.g.\ $K_{j'}$ or 
one of the incoming hadron momenta $P_i$ (or any linear fixed combination of 
them provided $K_\Lambda{\cdot}n$ is large), one ensures that 
the observed momentum fraction $z$ 
and intrinsic transverse momentum squared $k_T^2$ 
are Lorentz invariant. Moreover, 
for any of these choices the same $z$ and $k_T^2$ result, up to power 
corrections, which are neglected in our treatment. 
Below we will choose $n\,{=}\,K_{j'}$, for which 
$\sigma\,{=}\,(K_j{\cdot}K_\Lambda-z^{-1}M_\Lambda^2)/K_\Lambda{\cdot}K_{j'}$ 
is always small, 
of order hadronic scale squared divided by the partonic center of mass 
energy squared. The momentum fraction and intrinsic transverse momentum of the
outgoing parton are then 
\begin{equation}
z=\frac{K_\Lambda{\cdot}K_{j'}}{K_j{\cdot}K_{j'}}\ ,\qquad\text{and}\qquad
k_T = K_j-z^{-1}K_\Lambda\ .
\label{zandkt}
\end{equation}
Note that $k_T$ refers to the transverse component of the parton's momentum
w.r.t.\ the $\Lambda$ direction.
The transverse component $K_{\Lambda\,T}$ of the $\Lambda$ w.r.t.\ the jet
direction is $K_{\Lambda\,T}\,{=}\,{-}zk_T\,{=}\,K_\Lambda{-}zK_j$ (note that
the subscripts $T$ of $K_{\Lambda\,T}$ and $k_T$ refer to different frames
\cite{Mulders:1995dh}). 
The TMD fragmentation functions are commonly written with this vector as a
variable. 

For the cross section we take as a starting point the assumption that for each
observed hadron separately the soft physics factorizes from the hard physics: 
\begin{align}
d\sigma_{pp\rightarrow(\Lambda^\uparrow j)j' X}
&=\frac{1}{2s}\,\frac{d^3K_\Lambda}{(2\pi)^3\,2E_\Lambda}\,
\frac{d^3K_j}{(2\pi)^3\,2E_j}\,
\frac{d^3K_{j'}}{(2\pi)^3\,2E_{j'}}\,
2(2\pi)^3{\int}dx_1d^2p_{1T}\ dx_2d^2p_{2T}\;\ 
(2\pi)^4\delta^4(p_1{+}p_2{-}K_j{-}K_{j'})\nonumber\\
%%%%
&\mspace{80mu}\times\ 
\sum_{a,{\cdots},d}
\Phi_a(x_1{,}p_{1T})\,\otimes\,
\Phi_b(x_2{,}p_{2T})\,\otimes\,
\big|\,H^{ab\rightarrow cd}(p_1{,}p_2{,}K_j{,}K_{j'})\,\big|^2\,\otimes\,
\Delta_c(z{,}k_T)\ ,\label{crosssec}
\end{align}
with $s\,{\equiv}\,(P_1{+}P_2)^2\,{=}\,E_{\text{cm\,tot}}^2$. The convolutions
$\otimes$ indicate the appropriate Dirac and color traces for 
the partonic hard squared amplitude $|H|^2$. 
The momentum conservation delta function can furthermore be written as
\begin{equation}\begin{split}\label{DELTA}
\delta^4(p_1{+}p_2{-}&K_j{-}K_{j'})\\
&=\frac{2}{s}\,
\delta\big(x_1
-\tfrac{1}{2}(x_{j\perp}e^{\eta_j}{+}x_{j'\perp}e^{\eta_{j'}})\big)\,
\delta\big(x_2
-\tfrac{1}{2}(x_{j\perp}e^{-\eta_j}{+}x_{j'\perp}e^{-\eta_{j'}})\big)\,
\delta^2(p_{1T\perp}{+}p_{2T\perp}
{-}K_{j\perp}{-}K_{j'\perp})\ ,
\end{split}\end{equation}
where the $\eta_{j/j'}\,{=}\,{-}\ln\tan(\frac{1}{2}\theta_{j/j'})$ are the 
pseudorapidities of the outgoing jets and $\theta_{j/j'}$ denote polar
angles in the c.o.m.\ frame of the two incoming hadrons.
The $p_{iT\perp}$ indicate the projections of the intrinsic transverse momenta
$p_{iT}$ of the incoming partons in the plane perpendicular to the beam axis.
A factorization theorem as in Eq.\ (\ref{crosssec}) 
has not been proven rigorously for this process,
but we believe that it is plausible to assume that if it exists,
it will be of the schematic form of the expression in~Eq.\ (\ref{crosssec}),
possibly up to soft factors.

In principle there are several effects that contribute to the spin 
dependence of the cross section in Eq.\ (\ref{crosssec}), 
viz.\ the Boer-Mulders effect and the Sivers fragmentation effect.
In the Boer-Mulders mechanism \cite{BM98} the single spin asymmetry 
arises as a consequence of a correlation between the transverse momentum 
and transverse spin of a quark inside an unpolarized initial hadron.
Hence, it relies on a $k_T$-effect in the initial 
state that will have to translate into an asymmetry in the final-state.
Through the hard partonic scattering there can be $k_T$-smearing, leading to a
dilution of the asymmetry in the final state. We will, therefore, 
assume that the Boer-Mulders effect is less important than the Sivers 
fragmentation effect. 
The latter is a $k_T$-effect that is directly in the 
final state, such that there is no dilution from $k_T$-smearing in the partonic
scattering. In addition, one may expect that the chiral-odd 
contributions ($\propto h_1^\perp H_1$) are generally smaller than the 
chiral-even ones ($\propto f_1 D_{1T}^\perp$),
similar to what was found in Ref.\ \cite{Bacchetta:2007sz} for the process 
$p^\uparrow \, p \to \gamma \, \text{jet} \, X$ using positivity bounds. 
Based on these considerations, we will simply neglect the intrinsic transverse
momenta $p_{iT}$ of the incoming particles, such that only the Sivers
fragmentation mechanism contributes. As mentioned in the introduction, this
effect is described by the TMD fragmentation function 
$D_{1T}^\perp(z{,}K_{\Lambda\,T}^2)$, 
which enters the expression in~Eq.\ (\ref{crosssec}) 
through the parameterization of the fragmentation 
correlator~\cite{Mulders:1995dh,Bacchetta:2004jz,Bacchetta:2006tn}
(suppressing Wilson lines and with the convention $\epsilon^{0123}\,{=}\,{+}1$)
\begin{subequations}\label{FragCor}
\begin{align}
\Delta(z{,}k_T{;}K_\Lambda,S_\Lambda)
&=\sum_X\frac{1}{z}{\int}\frac{d(\xi{\cdot}K_\Lambda)\,d^2\xi_T}{(2\pi)^3}\ 
e^{-ik\cdot\xi}
\langle0|\,\psi(0)\,|\Lambda{;}X\rangle
\langle\Lambda{;}X|\,\overline\psi(\xi)\,|0\rangle\,\big\rfloor_{\xi\cdot n=0}
\label{FragCorA}\displaybreak[0]\\
%%%%
&=\Big(\,D_1(z{,}K_{\Lambda\,T}^2)
-D_{1T}^\perp(z{,}K_{\Lambda\,T}^2)
\frac{\epsilon^{\mu\nu\rho\sigma}K_{\Lambda\,\mu}
k_\nu S_{\Lambda\,\rho} n_\sigma}
{M_\Lambda\,(K_\Lambda{\cdot}n)}\,\Big)\,\slash K_\Lambda
+\text{other functions}\label{FragCorB}\displaybreak[0]\\
%%%%
&=\Big(\,D_1(z{,}K_{\Lambda\,T}^2)
+D_{1T}^\perp(z{,}K_{\Lambda\,T}^2)
\frac{(\hat{\boldsymbol k}{\times}\boldsymbol K_{\Lambda}){\cdot}
\boldsymbol  S_\Lambda}{z\,M_\Lambda}\,\Big)\,\slash K_\Lambda
+\text{other functions}\ ,\label{FragCorC}
\end{align}
\end{subequations}
in the case of quark fragmentation. Here $k$ denotes the quark momentum, which
equals the jet momentum $K_j$, and $n$ is the Sudakov vector discussed before. 
The last line (\ref{FragCorC}) 
holds in any reference frame where $\boldsymbol n$ and
$\boldsymbol{\hat k}$ point in opposite directions.  
Sometimes the notation 
$\Delta^ND_{\Lambda^\uparrow/a}\,{=}\,|\boldsymbol K_{\Lambda\,T}|
D_{1T}^{a\,\perp}/zM_\Lambda$ is used to indicate the Sivers
fragmentation function \cite{ABDM01}.

As mentioned in the introduction, there is a complication specifically 
associated with TMD fragmentation functions. 
The Sivers fragmentation function of Eqs.~\eqref{FragCorB}
and~\eqref{FragCorC} is actually a combination of the two functions in 
Fig.~\ref{ToddFragmentation}, arising from two different types of
interactions. In the hadronic cross section the function 
$D_{1T}^\perp(z{,}K_{h\,T}^2)$ generated by the final-state interactions will 
appear convoluted with the usual partonic scattering cross sections.
However, due to the dependence of the soft gluon interactions on the
partonic subprocesses (in particular on its color flow structure), 
the TMD fragmentation function $\widetilde D{}_{1T}^\perp(z{,}K_{h\,T}^2)$
generated by this effect will appear convoluted with modified partonic
scattering cross sections.
The contribution of the Sivers fragmentation mechanism to the hadronic cross 
section will thus have the form
\begin{equation}
\frac{d\hat\sigma_{ab\rightarrow cd}}{d\hat t}\,D_{1T}^{c\,\perp}(z{,}K_{h\,T}^2)
+\frac{d\hat\sigma_{ab\rightarrow[c]d}}{d\hat t}\,
\widetilde D_{1T}^{c\,\perp}(z{,}K_{h\,T}^2)\ ,
\end{equation}
where the $d\hat\sigma_{ab\rightarrow cd}$ are the usual partonic scattering
cross sections and the $d\hat\sigma_{ab\rightarrow[c]d}$ are the modified
partonic scattering cross sections (referred to as 
gluonic pole cross sections\footnote{For $T$-odd gluon correlators in
  principle two distinct gluonic pole cross sections 
$d\hat\sigma^{(f)}$ and $d\hat\sigma^{(d)}$ appear. However, 
the latter cross section vanishes for the gluon-gluon scattering 
contribution~\cite{Bomhof:2006ra}, the only contribution considered in Sec.\
\ref{phenom}. 
Therefore, the function $\widetilde D_{1T}^{g\,\perp}$ in this paper refers to
$\widetilde D_{1T}^{g\,\perp\,(f)}$.}) 
calculated in Refs.~\cite{Bacchetta:2005rm,Bomhof:2006ra}.
Therefore, the Sivers fragmentation functions in Eqs.~\eqref{FragCorB}
and~\eqref{FragCorC} should be read as $D_{1T}^\perp\,{\rightarrow}\,
D_{1T}^\perp{+}C_G^{[D]}\widetilde D{}_{1T}^\perp$,
where the $C_G^{[D]}$ are color fractions that depend on the particular term
in the partonic squared amplitude $H^*H$ which is given by
the cut Feynman diagram $D$. 
These factors are perturbatively calculable and, 
once absorbed in the partonic scattering cross sections,
lead to the gluonic pole cross sections. 
The gluonic pole cross sections, then, 
are gauge invariant weighted sums of cut Feynman diagrams $D$ 
with the color fractions $C_G^{[D]}$ as weight factors. 

After neglecting intrinsic transverse momentum 
effects for the incoming particles the Sivers
fragmentation contribution to the cross section Eq.~\eqref{crosssec} 
{\em in the jet-jet c.o.m.-frame\/} becomes 
\begin{gather}
\frac{E_\Lambda\,E_j\,E_{j'}\,d\sigma_{pp\rightarrow(\Lambda^\uparrow j)j' X}}
{d^3K_{\Lambda}\,d^3K_j\,d^3K_{j'}}
=\frac{1}{\pi}\delta^2(K_{j\perp}{+}K_{j'\perp})
\sum_{a,{\cdots},d}
x_1f_1^a(x_1)\,x_2f_1^b(x_2)
\bigg\{\,\frac{d\hat\sigma_{ab\rightarrow cd}}{d\hat t}\,
zD_1^c(z{,}(K_\Lambda{-}zK_j)^2)\nonumber\\
%%%%
\mspace{370mu}
+\frac{\boldsymbol S_\Lambda{\cdot}
(\boldsymbol {\hat K}{}_j{\times}\boldsymbol K_\Lambda)}{z\,M_\Lambda}
\bigg[\,\frac{d\hat\sigma_{ab\rightarrow cd}}{d\hat t}\,
zD_{1T}^{c\,\perp}(z{,}(K_\Lambda{-}zK_j)^2)\nonumber\\
%%%%
\mspace{520mu}
+\frac{d\hat\sigma_{ab\rightarrow[c]d}}{d\hat t}\,
z\widetilde D{}_{1T}^{c\,\perp}(z{,}(K_\Lambda{-}zK_j)^2)\,\bigg]\,\bigg\}\ ,
\label{CS1}
\end{gather}
where the momentum fractions $x_1$ and $x_2$ have been fixed by the first two
delta functions on the r.h.s.\ of Eq.~\eqref{DELTA}. 
The partonic scattering cross sections $d\hat\sigma_{ab\rightarrow cd}$ and the 
gluonic pole cross sections $d\hat\sigma_{ab\rightarrow[c]d}$ are functions of the Mandelstam variables $\hat s\,{\equiv}\,(p_1{+}p_2)^2$, 
$\hat t\,{\equiv}\,(k{-}p_1)^2$ and $\hat u\,{\equiv}\,(k'{-}p_1)^2$ or, alternatively, of $\hat s$ and the variable $y$
defined through 
\begin{gather}
y\equiv-\frac{\hat t}{\hat s}=\frac{1}{e^{\eta_j-\eta_{j'}}{+}1}\ ,\quad
%%%%
\mbox{and} \quad
-\frac{\hat u}{\hat s}=1-y\ .
\label{ydef}
\end{gather}
Note that the expression of $y$ in terms of $\eta_j$ and $\eta_{j'}$ in Eq.~\eqref{ydef} holds in the center of mass frame of the initial hadrons. Since we are ignoring
masses w.r.t.\ $|\boldsymbol K_{j\perp}|$, pseudorapidity and rapidity are
the same and since differences of rapidities are invariant under boosts, 
Eq.\ (\ref{ydef}) also holds in the jet-jet c.o.m.-frame. It should be
emphasized that Eq.\ (\ref{CS1}) does not hold in the c.o.m.-frame of the
incoming hadrons. 

Eq.\ (\ref{CS1}) forms the central result of this paper. Any nonzero 
measurement of the $\boldsymbol S_\Lambda{\cdot}
(\boldsymbol {\hat K}{}_j{\times}\boldsymbol K_\Lambda)$ asymmetry 
implies a nonzero Sivers effect in the fragmentation process and will thus be 
very interesting. It will imply nonzero $D_{1T}^\perp$ and/or 
$\widetilde D{}_{1T}^\perp$. Due to the different partonic cross sections
appearing in
front of these two different Sivers fragmentation functions, measuring the $y$
dependence of the asymmetry in principle offers the possibility to separate 
the two contributions, 
and to learn about their relative importance. This we will
explore further in the next section. 

If one were to study instead the process $p(P_1){+}p(P_2)
{\rightarrow}\left(\Lambda^\uparrow(K_\Lambda)\text{jet}(K_j)\right){+}X$ 
without constructing $K_{j'}$, then one does not have the full partonic 
kinematic information and one will be probing the combination
\begin{equation}
D_{1T}^{c\,\perp}(z{,}K_{h\,T}^2)
+\frac{d\hat\sigma_{ab\rightarrow[c]d}/d\hat t}{d\hat\sigma_{ab\rightarrow
    cd}/d\hat t}\,\, \widetilde D_{1T}^{c\,\perp}(z{,}K_{h\,T}^2)\ ,
\end{equation}
averaged over $y$. In addition, referring to a correlation 
$\boldsymbol S_\Lambda{\cdot}
(\boldsymbol {\hat K}{}_j{\times}\boldsymbol K_\Lambda)$
in the jet-jet
c.o.m.-frame, involving three-vectors (corresponding to the choice $n \propto 
K_{j'}$) would not make sense. One could for instance choose $n \propto P_1$ 
and consider this correlation in terms of three-vectors in any frame in which 
$\boldsymbol{\hat{P}}_1$ and $\boldsymbol{\hat{K}}_j$ point in opposite 
directions. In the remainder of
this article we will focus on the case where $K_{j'}$ is measured and 
present results in the jet-jet c.o.m.-frame. 

\section{Phenomenology\label{phenom}}

One can decompose the hadronic cross section in a spin-independent and 
a spin-dependent part:
\begin{equation}
\frac{d\sigma}{d^3K_{\Lambda}\,d^3K_j\,d^3K_{j'}}
=\frac{d\sigma_U}{d^3K_{\Lambda}\,d^3K_j\,d^3K_{j'}}
+\frac{\boldsymbol S_\Lambda{\cdot}
(\boldsymbol {\hat K}{}_j{\times}\boldsymbol K_\Lambda)}{z\,M_\Lambda}
\frac{d\sigma_T}{d^3K_{\Lambda}\,d^3K_j\,d^3K_{j'}}\ .
\end{equation}
Comparing to Eq.~\eqref{CS1} 
it follows that within our approximations the ratio of the spin-dependent and spin-independent cross sections is given by
\begin{align}
\frac{d\sigma_T}{d\sigma_U}
&=\frac{\sum\limits_{a,{\cdots},d}x_1f_1^a(x_1)\,x_2f_1^b(x_2)\,
\big\{\,d\hat\sigma_{ab\rightarrow cd}\,
D_{1T}^{c\,\perp}\big(z{,}(K_\Lambda{-}zK_j)^2)
+d\hat\sigma_{ab\rightarrow[c]d}\,
\widetilde D{}_{1T}^{c\,\perp}\big(z{,}(K_\Lambda{-}zK_j)^2)\,\big\}}
{\sum\limits_{a,{\cdots},d}
x_1f_1^a(x_1)\,x_2f_1^b(x_2)\,d\hat\sigma_{ab\rightarrow cd}\,
D_1^c\big(z{,}(K_\Lambda{-}zK_j)^2)}\ .
\label{ratio}
\end{align}
This ratio appears in the single spin asymmetry
\begin{align}
\text{SSA}
&=\frac{d\sigma({+}\boldsymbol S_\Lambda)\,
{-}\,d\sigma({-}\boldsymbol S_\Lambda)}
{d\sigma({+}\boldsymbol S_\Lambda)\,
{+}\,d\sigma({-}\boldsymbol S_\Lambda)}
=\frac{\boldsymbol S_\Lambda{\cdot}
(\boldsymbol {\hat K}{}_j{\times}\boldsymbol K_\Lambda)}{z\,M_\Lambda}\,
\frac{d\sigma_T}{d\sigma_U}\ .
\end{align}

\begin{figure}
\centering
\begin{minipage}[t]{5cm}
\centering
\psfrag{y}[cc][cc]{$y$}
\includegraphics[width=\textwidth]{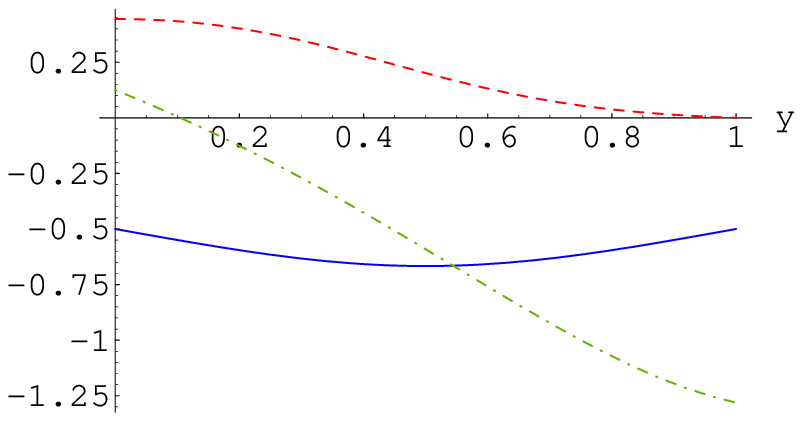}
(a)
\end{minipage}\hspace{1.5cm}
\begin{minipage}[t]{5cm}
\centering
\psfrag{eta}[cc][cc]{$\eta_j$}
\includegraphics[width=\textwidth]{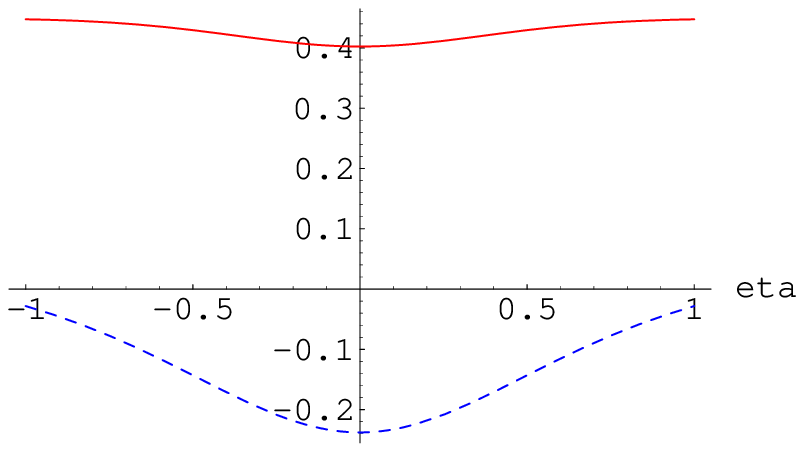}
(b)
\end{minipage}
\parbox{0.95\textwidth}{\caption{
(a) Solid line: $a_g(y)$; Dash-dotted line: $a_q(y)$;
Dashed line: $b(y)$.
(b) The combinations $b(y){+}b(1{-}y)$ (solid line) and 
$b(y)a_q(y){+}b(1{-}y)a_q(1{-}y)$ (dashed line) displayed as functions of 
$\eta_j$ for $\eta_j=-\eta_{j'}$ events. All graphs are for $N\,{=}\,3$.
\label{abfuncts}}}
\end{figure}

The relevant partonic hard functions for our study are~\cite{Bomhof:2006ra}
\begin{subequations}
\begin{alignat}{2}
&\frac{d\hat\sigma_{gg\rightarrow gg}}{d\hat t}
=\frac{\pi\alpha_s^2}{\hat s^2}\frac{2N^2}{N^2{-}1}
\frac{(\hat s^4{+}\hat t^4{+}\hat u^4)(\hat s^2{+}\hat t^2{+}\hat u^2)}
{2\hat s^2\hat t^2\hat u^2}\ ,&\mspace{30mu}
%%%%
&\frac{d\hat\sigma_{gg\rightarrow[g]g}}{d\hat t}
=-\frac{\pi\alpha_s^2}{\hat s^2}\frac{2N^2}{N^2{-}1}
\frac{(\hat s^2{+}\hat t^2{+}\hat u^2)^2}{4\hat t^2\hat u^2}\ ,\\
%%%%
%%%%
&\frac{d\hat\sigma_{qg\rightarrow qg}}{d\hat t}
=-\frac{\pi\alpha_s^2}{\hat s^2}
\frac{\hat s^2{+}\hat u^2}{2\hat s\hat u}
\Big(\,\frac{\hat s^2{+}\hat u^2}{\hat t^2}{-}\frac{1}{N^2}\,\Big)\ ,&
%%%%
&\frac{d\hat\sigma_{qg\rightarrow[q]g}}{d\hat t}
=\frac{\pi\alpha_s^2}{\hat s^2}
\frac{\hat s^2{+}\hat u^2}{2\hat s\hat u}
\Big(\,\frac{\hat s^2}{\hat t^2}{-}
\frac{N^2{+}1}{N^2{-}1}\Big(\,\frac{\hat u^2}{\hat t^2}
{-}\frac{1}{N^2}\,\Big)\,\Big)\ .\mspace{-30mu}
\end{alignat}
\end{subequations}
All cross sections are given for massless partons. 
The cross sections with the initial or final state particles interchanged are
obtained from these by a $\hat t{\leftrightarrow}\hat u$ substitution.
We also define the ratios (displayed in Fig.~\ref{abfuncts}a)
\begin{subequations}
\begin{gather}
a_g(y)=\frac{d\hat\sigma_{gg\rightarrow[g]g}}
{d\hat\sigma_{gg\rightarrow gg}}
=-\frac{\hat s^2}{2}\,\frac{\hat s^2{+}\hat t^2{+}\hat u^2}
{\hat s^4{+}\hat t^4{+}\hat u^4}
=-\frac{1}{2}\,
\frac{1\,{+}\,y^2\,{+}\,(1{-}y)^2}{1\,{+}\,y^4\,{+}\,(1{-}y)^4}\ ,\\
%%%%
a_q(y)
=\frac{d\hat\sigma_{qg\rightarrow[q]g}}{d\hat\sigma_{qg\rightarrow qg}}
=\frac{N^2{+}1}{N^2{-}1}
-\frac{2N^2}{N^2{-}1}
\frac{\hat s^2}{\hat s^2{+}\hat u^2{-}\frac{1}{N^2}\hat t^2}
=\frac{N^2{+}1}{N^2{-}1}
-\frac{2N^2}{N^2{-}1}
\frac{1}{1{+}(1{-}y)^2{-}\frac{1}{N^2}y^2}\ ,\\
%%%%
%%%%
b(y)
=\frac{d\hat\sigma_{qg\rightarrow qg}}{d\hat\sigma_{gg\rightarrow gg}}
=-\frac{N^2{-}1}{2N^2}
\frac{\hat s\hat u
(\hat s^2{+}\hat u^2)(\,\hat s^2{+}\hat u^2{-}\frac{1}{N^2}\hat t^2\,)}
{(\hat s^4{+}\hat t^4{+}\hat u^4)(\hat s^2{+}\hat t^2{+}\hat u^2)}
=\frac{N^2{-}1}{2N^2}
\frac{(1{-}y)(1{+}(1{-}y)^2)(\,1{+}(1{-}y)^2{-}\frac{1}{N^2}y^2\,)}
{(1{+}y^4{+}(1{-}y)^4)(1{+}y^2{+}(1{-}y)^2)}\ .
\end{gather}
\end{subequations}

\begin{figure}
\centering
\psfrag{xperp}[cc][cc]{$x_\perp$}
\begin{minipage}[t]{6cm}
\centering
\includegraphics[width=\textwidth]{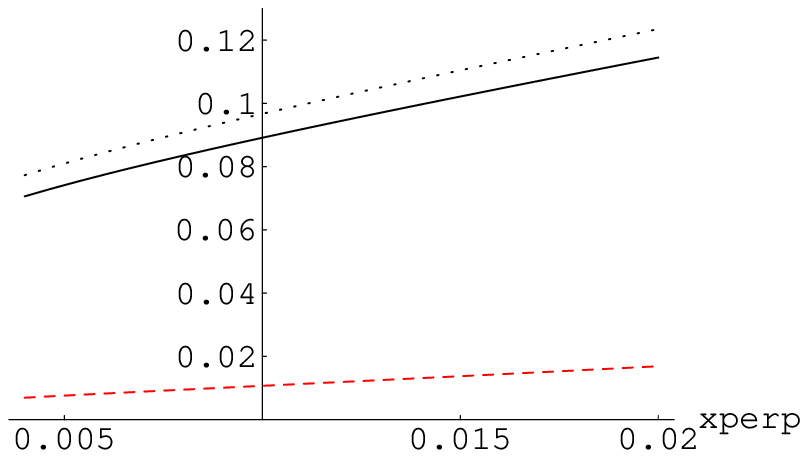}
(a)
\end{minipage}\hspace{1cm}
%%%%
\begin{minipage}[t]{6cm}
\centering
\includegraphics[width=\textwidth]{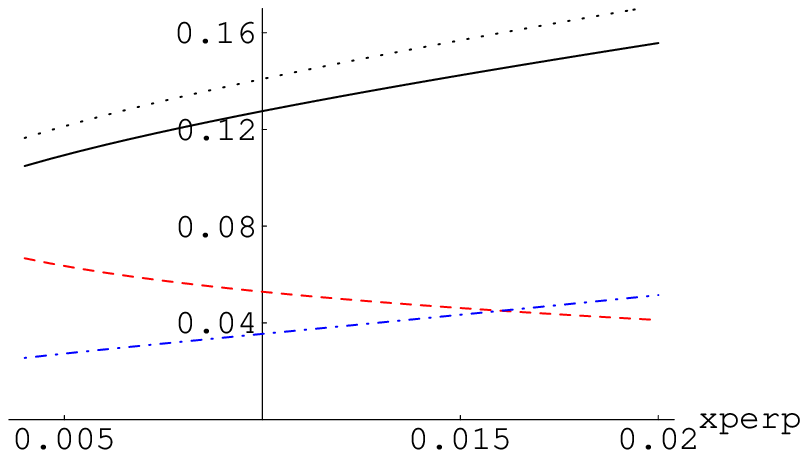}
(b)
\end{minipage}
\parbox{0.95\textwidth}{\caption{
(a) The partonic cross sections multiplied by the unpolarized collinear distribution functions, 
integrated over ${-}1\,{\leq}\,\eta_{j/j'}{\leq}\,1$, normalized to the 
corresponding quantity for gluon-gluon scattering
(numerator and denominator integrated separately). 
Solid line: $qg{\rightarrow}qg/gg{\rightarrow}gg$,
dashed line: $qq'{\rightarrow}qq'/gg{\rightarrow}gg$.
For the unpolarized distribution functions we have used the CTEQ5L
distribution functions evaluated at 10~GeV.
For comparison $qg{\rightarrow}qg/gg{\rightarrow}gg$ at 20~GeV was also included (dotted line).
We have approximated $x_{j\perp}\,{\approx}\,x_{j'\perp}\,{\equiv}\,x_\perp$.
(b) As in (a), but now with 
$qq'{\rightarrow}qq'/qg{\rightarrow}qg$ (solid line),
$gg{\rightarrow}q\bar q/qg{\rightarrow}qg$ (dashed line) and 
$qq{\rightarrow}qq/qg{\rightarrow}qg$ (dash-dotted line). 
Over the entire kinematical region considered here all other partonic channels are dominated by those shown in the figure.
For comparison $qq'{\rightarrow}qq'/gg{\rightarrow}gg$ at 20~GeV was also included (dotted line).
\label{comparison}}}
\end{figure}

In Fig.~\ref{comparison} we show the expressions 
$x_1f_1^a(x_1)x_2f_1^b(x_2)d\hat\sigma_{ab\rightarrow cd}$ appearing in the
denominator of Eq.\ (\ref{ratio}), integrated over the pseudorapidity ranges 
${-}1\,{\leq}\,\eta_{j,j'}\,{\leq}\,{+}1$, normalized to the corresponding
quantity for gluon-gluon scattering (Fig.~\ref{comparison}a) and quark-gluon
scattering (Fig.~\ref{comparison}b). 
The pseudorapidity integration range is inspired by the $\eta$ coverage of
the ALICE detector at LHC, which has the particle identification 
capabilities that allow
to measure $\Lambda$'s with transverse momenta $p_T^{}$ of several GeV$/c$. 
We have taken $\sqrt s\,{=}\,14$~TeV as is relevant to the LHC situation 
and for the unpolarized distribution functions we have used the CTEQ5L
parametrizations evaluated at the scale $10$~GeV. 
The figures do not change significantly with the scale in this region
(see Figure~\ref{comparison}).
What we infer from the figure is that the combination
$x_1f_1^a(x_1)x_2f_1^b(x_2)d\hat\sigma_{ab\rightarrow cd}$ which determines the
inclusive cross section 
is dominated by the gluon-gluon ($gg{\rightarrow}gg$) scattering process in
this particular kinematic region.
The quark-gluon ($qg{\rightarrow}qg$) scattering contribution is on the order
of $10\%$ of the ($gg{\rightarrow}gg$) contribution at perpendicular scales
$x_\perp$ typical for LHC, if we consider jets with transverse momenta 
between $30$ and $100$ GeV$/c$. 
Similarly, the other partonic contributions are of the order of $10\%$ 
(or less) of the $qg{\rightarrow}qg$ contribution (Fig.~\ref{comparison}b).

\begin{figure}
\centering
\psfrag{xperp}[lc][lc]{$\ \ x_\perp$}
\begin{minipage}[t]{7cm}
\centering
\includegraphics[width=\textwidth]{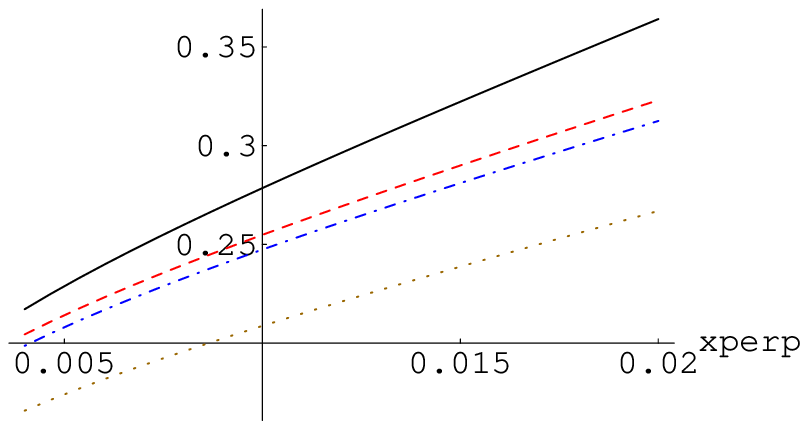}
(a)
\end{minipage}\hspace{5mm}
%%%%
\begin{minipage}[t]{7cm}
\centering
\includegraphics[width=\textwidth]{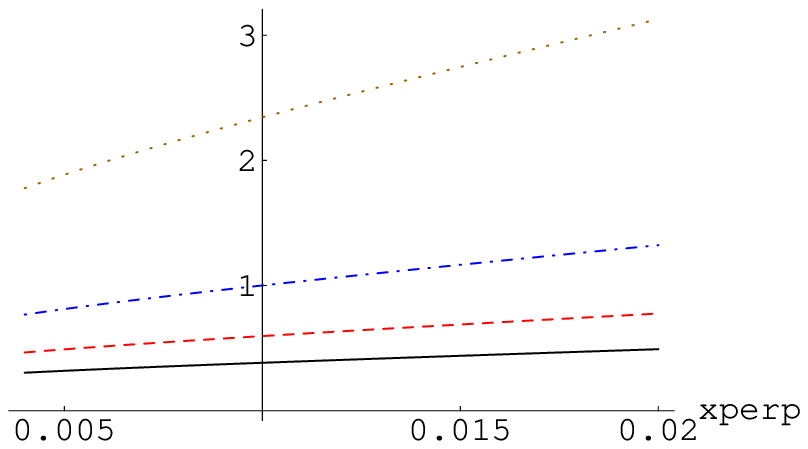}
(b)
\end{minipage}
\parbox{0.95\textwidth}{\caption{
Expression in Eq.~\eqref{uitdrukking} integrated over 
${-}1\,{\leq}\,\eta_{j/j'}{\leq}\,1$ (numerator and denominator integrated 
separately) with (a) AKK and (b) FSV $\Lambda$-fragmentation 
functions, for
$z\,{=}\,0.1$ (solid line),
$z\,{=}\,0.3$ (dashed line),
$z\,{=}\,0.5$ (dash-dotted line) and
$z\,{=}\,0.7$ (dotted line).
Distribution and fragmentation functions are evaluated at $10$~GeV.
\label{AKKvsFSV}}}
\end{figure}

Whether or not the unpolarized $\Lambda$ production 
cross section $d\sigma_U$ can also be 
approximated by the gluon-gluon scattering contribution, depends heavily on 
the $\Lambda$ fragmentation functions employed. 
Several such functions have been fitted
to experimental data \cite{FSV,Indumathi:1998am,Boros:2000ex,AKK},
unfortunately with vastly different results for the ratios $D_1^g/D_1^q$ for
the various quark and antiquark flavors. If one restricts to the subprocesses
$gg{\rightarrow}gg$ and $qg{\rightarrow}qg$, then the unpolarized cross section
$d\sigma_U$ contains
\begin{align}
d\sigma_{U}
&\sim
x_1f_1^g(x_1)\,x_2f_1^g(x_2)\,d\hat\sigma_{gg\rightarrow gg}\,D_1^g(z)\,
\big\{\,1+\varepsilon(\eta_1{,}\eta_2)\,\big\}\ ,
\end{align}
with
\begin{subequations}
\begin{align}
\varepsilon
& \equiv \frac{\sum_qx_1f_1^q(x_1)\,x_2f_1^g(x_2)\,\left[\,
d\hat\sigma_{qg\rightarrow qg}\,D_1^q(z)
{+}d\hat\sigma_{qg\rightarrow gq}\,D_1^g(z)\,\right]}
{x_1f_1^g(x_1)\,x_2f_1^g(x_2)\,d\hat\sigma_{gg\rightarrow gg}\,D_1^g(z)}
\nonumber\\
&\mspace{200mu}
+\frac{\sum_q\,x_1f_1^g(x_1)\,x_2f_1^q(x_2)\,\left[\,
d\hat\sigma_{gq\rightarrow qg}\,D_1^q(z)
{+}d\hat\sigma_{gq\rightarrow gq}\,D_1^g(z)\,\right]}
{x_1f_1^g(x_1)\,x_2f_1^g(x_2)\,d\hat\sigma_{gg\rightarrow gg}\,D_1^g(z)}
\label{uitdrukking}\\
%%%%
&=b(y)\frac{\sum_qf_1^q(x_1)D_1^q(z)}{f_1^g(x_1)D_1^g(z)}
+b(1{-}y)\frac{\sum_qf_1^q(x_1)}{f_1^g(x_1)}
+b(y)\frac{\sum_qf_1^q(x_2)}{f_1^g(x_2)}
+b(1{-}y)\frac{\sum_qf_1^q(x_2)D_1^q(z)}{f_1^g(x_2)D_1^g(z)}\ .
\end{align}
\end{subequations}
The function $\varepsilon$ is quite sensitive to the choice of unpolarized
fragmentation functions, as demonstrated in Fig.~\ref{AKKvsFSV}, 
which shows a comparison using the functions by De Florian et al.~\cite{FSV} (FSV) and 
by Albino et al.~\cite{AKK} (AKK), for a factorization scale of 10 GeV. 

For the AKK set the assumption of gluon-gluon scattering dominance may be good
to 30\%, but for the FSV set one can reach this conclusion 
only for small $z$ values. At ALICE $\Lambda$'s with transverse momentum of
several GeV are expected to be measured in sufficiently large quantities. A
typical $z$ value could be $0.1$ for a 5 GeV $\Lambda$ within a 50 GeV jet.
Therefore, we expect the midrapidity data to be restricted to 
small $z$ values at LHC and the assumption of gluon-gluon scattering 
dominance to be valid despite the large variation in $\Lambda$ 
fragmentation function sets. 

A future investigation of
$\Lambda$ production at LHC or RHIC should, therefore, 
first focus on describing the
unpolarized cross section correctly, especially regarding the $y$-dependence,
before the $y$-dependence of the polarization effect is studied. Therefore, 
even if no polarization effect is seen, the unpolarized data obtained by the
proposed measurement can in any case lead to a considerable improvement in 
the determination of the unpolarized $\Lambda$ fragmentation functions.   

To illustrate the main idea regarding the $y$ dependence of the asymmetry, 
we will simply assume that 
the unpolarized cross section is known sufficiently
well and that gluon-gluon scattering dominates, and 
consider two distinct scenarios: 1) 
 $D_{1T}^{g\,\perp}\,{\neq}\,0$; 2) $D_{1T}^{g\,\perp}\,{=}\,0$ or more
 generally, $D_{1T}^{g\,\perp}\,{\ll}\,D_{1T}^{q\,\perp}$. 
   
In scenario 1 ($D_{1T}^{g\,\perp}\,{\neq}\,0$)
 one finds from Eq.\ (\ref{ratio}): 
\begin{align}
\frac{d\sigma_T}{d\sigma_U}
&\approx
\frac{x_1f_1^g(x_1)\,x_2f_1^g(x_2)\,
d\hat\sigma_{gg\rightarrow gg}\,D_{1T}^{g\,\perp}(z{,}K_{\Lambda\,T}^2)
+x_1f_1^g(x_1)\,x_2f_1^g(x_2)\,d\hat\sigma_{gg\rightarrow[g]g}\,
\widetilde D{}_{1T}^{g\,\perp}(z{,}K_{\Lambda\,T}^2)}
{x_1f_1^g(x_1)\,x_2f_1^g(x_2)\,
d\hat\sigma_{gg\rightarrow gg}\,D_1^g(z{,}K_{\Lambda\,T}^2)}\nonumber\\
%%%%
&=\frac{D_{1T}^{g\,\perp}(z{,}K_{\Lambda\,T}^2)}{D_1^g(z{,}K_{\Lambda\,T}^2)}
+a_g(y)
\frac{\widetilde D{}_{1T}^{g\,\perp}(z{,}K_{\Lambda\,T}^2)}
{D_1^g(z{,}K_{\Lambda\,T}^2)}\ .
\end{align}
If there would be no Wilson line or color flow dependence in the fragmentation
process in the way as has been claimed in 
Refs.\ \cite{Metz:2002iz,Collins:2004nx}, 
$\widetilde D{}_{1T}^{g\,\perp}$ would arise with the usual partonic cross
section. In that case there is no need to consider two separate functions to
begin with and one should not find any $y$ dependence at all. 
Observing the $y$ dependence
$a_g(y)$ (which is rather weak unfortunately) would be a signal for 
color flow dependence
of the $T$-odd effects in the fragmentation process.   

In scenario 2 ($D_{1T}^{g\,\perp}\,{\ll}\,D_{1T}^{q\,\perp}$) one finds on the
other hand: 
\begin{align}
\frac{d\sigma_T}{d\sigma_U}
&\approx
\frac{\sum_q\big(\,x_1f_1^q(x_1)\,x_2f_1^g(x_2)\,
d\hat\sigma_{qg\rightarrow qg}\,D_{1T}^{q\,\perp}(z{,}K_{\Lambda\,T}^2)
+x_1f_1^q(x_1)\,x_2f_1^g(x_2)\,d\hat\sigma_{qg\rightarrow[q]g}\,
\widetilde D{}_{1T}^{q\,\perp}(z{,}K_{\Lambda\,T}^2)\,\big)}
{x_1f_1^g(x_1)\,x_2f_1^g(x_2)\,
d\hat\sigma_{gg\rightarrow gg}\,D_1^g(z{,}K_{\Lambda\,T}^2)}\nonumber\\
%%%%
&\mspace{80mu}
+\frac{\sum_q\big(\,x_1f_1^g(x_1)\,x_2f_1^q(x_2)\,
d\hat\sigma_{gq\rightarrow qg}\,D_{1T}^{q\,\perp}(z{,}K_{\Lambda\,T}^2)
+x_1f_1^g(x_1)\,x_2f_1^q(x_2)\,d\hat\sigma_{gq\rightarrow[q]g}\,
\widetilde D{}_{1T}^{q\,\perp}(z{,}K_{\Lambda\,T}^2)\,\big)}
{x_1f_1^g(x_1)\,x_2f_1^g(x_2)\,
d\hat\sigma_{gg\rightarrow gg}\,D_1^g(z{,}K_{\Lambda\,T}^2)}
\displaybreak[0]\nonumber\\
%%%%
&=
b(y)\,\frac{\sum_qf_1^q(x_1)D_{1T}^{q\,\perp}(z{,}K_{\Lambda\,T}^2)}
{f_1^g(x_1)D_1^g(z{,}K_{\Lambda\,T}^2)}
+b(y)a_q(y)\,
\frac{\sum_qf_1^q(x_1)\widetilde D{}_{1T}^{q\,\perp}(z{,}K_{\Lambda\,T}^2)}
{f_1^g(x_1)D_1^g(z{,}K_{\Lambda\,T}^2)}\nonumber\\
%%%%
&\mspace{80mu}
+b(1{-}y)\,
\frac{\sum_qf_1^q(x_2)D_{1T}^{q\,\perp}(z{,}K_{\Lambda\,T}^2)}
{f_1^g(x_2)D_1^g(z{,}K_{\Lambda\,T}^2)}
+b(1{-}y)a_q(1{-}y)\,
\frac{\sum_qf_1^q(x_2)\widetilde D{}_{1T}^{q\,\perp}(z{,}K_{\Lambda\,T}^2)}
{f_1^g(x_2)D_1^g(z{,}K_{\Lambda\,T}^2)}\ .%\nonumber\\
\end{align}
This expression has a rather involved $y$ dependence. It is probably too 
difficult to disentangle the separate contributions. Therefore, we suggest a
further step. Since the momentum fractions $x_1$ and $x_2$ are fixed by the 
jet momenta, one can select those events where they are equal,
which corresponds to $\eta_{j'}{=}\,{-}\eta_j$.
In that case the expression simplifies
\begin{align}
\frac{d\sigma_T}{d\sigma_U}
&\approx
\frac{\sum_qf_1^q(x_1)\left\{\,\left[\,b(y){+}b(1{-}y)\,\right]
D_{1T}^{q\,\perp}(z{,}K_{\Lambda\,T}^2)
+\left[\,b(y)a_q(y){+}b(1{-}y)a_q(1{-}y)\,\right]
\widetilde D{}_{1T}^{q\,\perp}(z{,}K_{\Lambda\,T}^2)\,\right\}}
{f_1^g(x_1)D_1^g(z{,}K_{\Lambda\,T}^2)}\ ,
\label{ratio2}
\end{align}
where now $y\,{=}\,(e^{2\eta_j}{+}1)^{-1}$ and 
$x_1\,{=}\,x_2\,{=}\,x_\perp/2\sqrt{y(1{-}y)}$. The two different combinations
of $y$-dependent terms are depicted in Fig.~\ref{abfuncts}b 
as a function of $\eta_j$. One can see that one term in Eq.\ (\ref{ratio2}) 
varies more strongly with
$\eta_j$ than the other term. This may possibly allow a discrimination of 
the two effects. 

\section{Discussion and conclusions}

In this paper we have proposed a measurement of $\Lambda$ polarization in the
process $p \, p \to \left(
\Lambda^\uparrow \text{jet}\right) \, \text{jet} \, X$, where the $\Lambda$ 
is part of one of the two observed jets which 
are almost back-to-back in the plane perpendicular to the beam axis. Unlike
the traditional measurement in the process $pp\to \Lambda^{\uparrow}X$, this
new observable need not vanish at midrapidity. This makes it of interest for
high energy collider experiments, such as RHIC, Tevatron and LHC, as they
typically can detect $\Lambda$'s at midrapidity. We have studied the asymmetry 
for the ALICE experiment specifically, due to the fact that ALICE has good 
PID capabilities that allow detection of $\Lambda$'s of momenta of 
several GeV/$c$. The experimental situation at LHC furthermore leads to the
dominance of gluon-gluon and gluon-quark scattering, allowing for a
simplification of the expressions to good approximation. 

Observation of a nonzero $\boldsymbol S_\Lambda{\cdot}
(\boldsymbol {\hat K}{}_j{\times}\boldsymbol K_\Lambda)$ asymmetry in the
jet-jet c.o.m.\ frame indicates a nonzero Sivers effect in the fragmentation
process. The $y$ dependence of this observable in principle could allow for 
a study of the relative importance of the two types of interactions that could
exhibit such a Sivers effect. It allows for a study of possible color-flow
dependence of the asymmetry and of the (non)universality of the Sivers 
fragmentation functions, issues currently of much interest and that cannot be
addressed in $pp\to \Lambda^{\uparrow}X$. But even if no
asymmetry is observed, the unpolarized data that is obtained from the
polarization measurement would help to constrain the unpolarized $\Lambda$
fragmentation functions, which are currently not well-determined. High energy
collider experiments such as to be performed at LHC would be very helpful
in this respect, as the question of factorization of the unpolarized cross
section would not pose a problem.   

The study of color flow dependence  
we suggest here can be done without an actual extraction of $D_{1T}^\perp$,
namely by just studying the dependence of the observable on particular
kinematic variables. This should prove useful even if a trustworthy extraction
of the Sivers 
fragmentation functions themselves turns out to be too difficult. 

Currently no knowledge on the magnitude of the various 
Sivers fragmentation functions is available, therefore, no predictions can be 
given for the actual size of the asymmetry. But
the $y$ dependence of the various contributing terms could be derived. They
were discussed for two different scenarios: one in which the gluon Sivers
fragmentation function is important to include and one in which it is
irrelevant compared to the quark function. 

The proposed measurement allows for a determination of the $z$ and $k_T^2$
dependence of the Sivers fragmentation functions. 
For completeness, we mention that there are other 
processes from which this information 
can be extracted. This can be done in electron-positron
annihilation experiments via the process $e^+ e^- \to \left(\Lambda^\uparrow
  \text{jet}\right)\, X$ 
in a straightforward manner \cite{BJM97}. It allows to 
probe the combination $D_{1T}^\perp + \widetilde D{}_{1T}^{\perp}$.
It can also be done via neutral or charged current semi-inclusive DIS: 
$\ell \, p \to \ell' \left(\Lambda^\uparrow \text{jet}\right)\, X$, where the 
combination $D_{1T}^\perp - \widetilde D{}_{1T}^{\perp}$ is probed. 
The comparison to the functions from $p \, p \to \left(
\Lambda^\uparrow \text{jet}\right) \, \text{jet} \, X$ 
would allow for another test of the (non)universality of the Sivers
fragmentation functions.

\begin{acknowledgments}
We thank Mauro Anselmino, Federico Antinori, Federico Carminati, 
Umberto D'Alesio, Do-Won Kim, Seyong Kim, 
Sungchul Lee and Werner Vogelsang 
for useful discussions.
This work was supported in part by the International Cooperation
Program of the KICOS (Korea Foundation for International Cooperation
of Science \& Technology).
The work of C.B.\ was supported by the foundation for Fundamental 
Research of Matter (FOM) and the National Organization for Scientific 
Research (NWO). 
\end{acknowledgments}

\end{document}